\documentclass[conference]{IEEEtran}
\IEEEoverridecommandlockouts
\usepackage{lipsum} % For dummy text

\usepackage{ifpdf}
\usepackage{cite}
\usepackage[cmex10]{amsmath}
\usepackage{amsfonts,amsthm}
\usepackage{array}
\usepackage{mdwmath}
\usepackage{mdwtab}
\usepackage{graphicx}
\usepackage{fixltx2e}
\usepackage{url}
\usepackage{fancyhdr}
\usepackage{lastpage}
\usepackage{xcolor}
\usepackage{amsmath,amssymb,amsfonts}
\usepackage{algorithmic, float}
\usepackage{graphicx}
\usepackage{textcomp}
\usepackage{amsthm}
\theoremstyle{definition}

\usepackage{enumitem}
\usepackage{subfig}
\usepackage{indentfirst}
\usepackage{comment}
\usepackage{caption}
\usepackage{multirow,tabularx}
\usepackage{xcolor}
\usepackage{mathtools}

\usepackage[linesnumbered,ruled,vlined]{algorithm2e}

\SetCommentSty{mycommfont}
\SetKwInput{KwInput}{Input}                % Set the Input
\SetKwInput{KwOutput}{Output}              % set the Output
\SetKwProg{myproc}{Procedure}{}{end}
\usepackage{eqparbox}

\usepackage{lipsum}
\usepackage{dblfloatfix} 
\usepackage{multirow}
\usepackage{caption}
\usepackage{amsthm}
\begin{document}

\title{%Empowering User Choice: Protecting Privacy in Time Series Data for Energy Supply Chain
%Empowering User Choice: Protecting Privacy in Time Series Data
Cooperative Local Differential Privacy: Securing Time Series Data in Distributed Environments }

\author{
\IEEEauthorblockN{
Bikash Chandra Singh\IEEEauthorrefmark{1}{\IEEEauthorrefmark{9}}, 
Md Jakir Hossain\IEEEauthorrefmark{2}\IEEEauthorrefmark{9}, 
Rafael Diaz\IEEEauthorrefmark{1}, 
Sandip~Roy\IEEEauthorrefmark{3,}\IEEEauthorrefmark{1},
Ravi Mukkamala\IEEEauthorrefmark{4},
Sachin~Shetty\IEEEauthorrefmark{3}
}
\IEEEauthorblockA{\IEEEauthorrefmark{1}School of Cybersecurity, Old Dominion University, Norfolk, VA 23529, USA \\ \{bsingh, rdiaz\}@odu.edu}

\IEEEauthorblockA{\IEEEauthorrefmark{2}Department of Computer Science and Engineering, Bangladesh University of Engineering and Technology, Dhaka \\ personal.jakirsab@gmail.com}

\IEEEauthorblockA{\IEEEauthorrefmark{4}Department of Computer Science, Old Dominion University, Norfolk, VA 23529, USA \\ rmukkama@odu.edu}

\IEEEauthorblockA{\IEEEauthorrefmark{3}Center for Secure \& Intelligent Critical Systems, Old Dominion University, Suffolk, VA 23435, USA \\ \{sroy, sshetty\}@odu.edu}
\thanks{\IEEEauthorrefmark{9}These authors contributed equally to this work.%
}%
}

\maketitle

\IEEEpeerreviewmaketitle

\begin{abstract}

%This paper addresses privacy concerns in time series data, which are increasingly collected through smart devices like mobile phones, wearables, and IoT sensors. The rapid growth of such data presents challenges for privacy preservation, especially when sensitive personal information is inadvertently exposed. Differential privacy (DP) has been widely adopted to safeguard privacy, but it assumes a trusted server, which is not always feasible. Local differential privacy (LDP) offers an alternative by allowing users to perturb their data before sharing, thus eliminating the need for a trusted intermediary. 

The rapid growth of smart devices—phones, wearables, IoT sensors, and connected vehicles—has led to an explosion of continuous time series data that offers valuable insights in healthcare, transportation, and more. However, this surge raises significant privacy concerns, as sensitive patterns can reveal personal details. While traditional differential privacy (DP) relies on trusted servers, local differential privacy (LDP) enables users to perturb their own data. However, traditional LDP methods perturb time series data by adding user-specific noise but exhibit vulnerabilities. For instance, noise applied within fixed time windows can be canceled during aggregation (e.g., averaging), enabling adversaries to infer individual statistics over time, thereby eroding privacy guarantees.

To address these issues, we introduce a Cooperative Local Differential Privacy (CLDP) mechanism that enhances privacy by distributing noise vectors across multiple users. In our approach, noise is collaboratively generated and assigned so that when all users' perturbed data is aggregated, the noise cancels out—preserving overall statistical properties while protecting individual privacy. This cooperative strategy not only counters vulnerabilities inherent in time-window-based methods but also scales effectively for large, real-time datasets, striking a better balance between data utility and privacy in multi-user environments.

%In this work, we introduce a novel Cooperative LDP (CLDP) method, where noise vectors are distributed among multiple users to ensure that aggregated data yields meaningful insights while individual privacy is maintained. Our approach utilizes structured noise generation via sine waves partitioned among users to facilitate noise cancellation during data aggregation. Furthermore, we propose an adaptive tossing strategy to fine-tune the privacy budget, striking a balance between noise application and aggregation accuracy. This method offers robust privacy protection while preserving the essential statistical characteristics of time series data, even under adversarial conditions.

\end{abstract}


\section{Introduction}

With the proliferation of smart devices like mobile phones, wearables, IoT sensors, and connected vehicles, data collection and processing have grown exponentially \cite{sisinni2018industrial, herath2022adoption}. This surge has given rise to continuous streams of time series data—ordered sets of values indexed by time—which are pivotal in many real-world applications, offering valuable insights across diverse domains \cite{deng2021compass}. For example, wearables analyze health metrics to provide personalized recommendations, while GPS services use location data for traffic management and route optimization.

Despite their transformative potential, these data streams raise critical privacy concerns \cite{allard2015chiaroscuro, zheng2021efficient}. Continuous data, such as heart rate patterns, can inadvertently expose sensitive details about an individual's daily life, making them susceptible to unauthorized surveillance. Addressing these privacy risks is crucial to fostering public trust and unlocking the full benefits of time series data for innovative and impactful applications.

To address privacy concerns in time series data, differential privacy (DP) has been widely adopted due to its strong privacy guarantees \cite{dong2019gaussian, yang2012differential, dwork2006calibrating}. However, DP assumes a trusted server to manage data, which is often unrealistic. Servers may unintentionally or intentionally compromise user privacy due to curiosity or commercial motives \cite{garfinkel2020randomness}. To mitigate this risk, local differential privacy (LDP) has emerged as a robust alternative, enabling users to perturb their data locally before sharing it, thus removing reliance on a trusted third party \cite{yang2023local}.

LDP methods for time series data typically follow two approaches: (i) selectively perturbing specific data points to preserve patterns or (ii) applying random perturbations across the dataset to prioritize privacy. While the first approach may compromise privacy in certain contexts—such as revealing power usage patterns in smart homes—the second can severely diminish data utility. For instance, random perturbations of heart rate data from wearables may obscure critical trends, affecting diagnoses and healthcare decisions. Moreover, many LDP-based methods struggle with scalability, especially for large datasets or real-time applications, as they often require identifying original data patterns before applying perturbations. 

Traditional LDP-based methods primarily focus on adding noise to perturb time series data for individual users. However, this approach leaves room for privacy vulnerabilities. For example, in methods that perturb data based on time windows, the added noise can often be canceled out during computations like summation or averaging over the time window. As a result, the statistical values for individual users may still be inferred, compromising their privacy over time.

For example, let suppose that a smart energy provider that collects minute-by-minute household power consumption data to forecast supply and demand in a region. To protect privacy, each household applies Local Differential Privacy (LDP) by injecting zero-mean random noise (e.g., $±2$ kWh) into their reported consumption values before transmitting data to the provider. However, when the provider aggregates this noisy data into hourly totals or averages (e.g., summing $60$ noisy data points), the random perturbations—designed to cancel out over time—lead to a near-accurate estimate of the true hourly consumption. For instance, if a household’s true usage is $0.5$ kWh/minute ($30$ kWh/hour), the LDP-noised values (e.g., $0.3, 0.7, 0.4$ kWh/minute) would still sum to $\approx30$ kWh/hour. Over time, this reveals behavioral patterns, such as when occupants wake up, use high-power appliances, or leave home—defeating the purpose of LDP.

This demonstrates how traditional LDP mechanisms, while privatizing individual data points, fail to safeguard against inference attacks on time-aggregated statistics, exposing sensitive long-term user habits.

To address this issue, we propose a Cooperative LDP (CLDP) method that enhances privacy by distributing noise vectors among multiple users. In this approach, noise vectors are collaboratively generated for \(n\) users and distributed in such a way that each user perturbs their data with a unique noise component. The users then share the perturbed data with the data collector. When the data collector aggregates the data (e.g., by summation) collected from the users, the perturbation noise is designed to cancel out, revealing only the aggregated statistical values without compromising the privacy of individual users.

This cooperative mechanism ensures that the data collector cannot infer specific statistical values for any single user while still allowing meaningful aggregated insights to be computed. By introducing a collective noise generation and distribution strategy, our method balances data utility and privacy protection more effectively, particularly for scenarios requiring secure aggregation across multiple users.

Therefore, the contributions of the paper are as follows:
\begin{itemize}
    \item We introduce a novel CLDP approach that enhances privacy by distributing noise vectors across multiple users. This ensures noise cancellation during data aggregation, preserving privacy while maintaining statistical insights.  
    
    \item The proposed method employs a cooperative strategy for generating and distributing noise, preventing the data collector from inferring individual user statistics while still enabling accurate aggregation.  

    \item  CLDP effectively balances privacy protection and data utility, allowing aggregated data to retain meaningful insights without compromising individual privacy.  

    \item Our method provides a robust framework for securely aggregating time series data across multiple users, ensuring strong privacy guarantees in adversarial settings and real-time applications.
\end{itemize}

The rest of the paper is structured as follows. Section \ref{section:preli} introduces key preliminaries, focusing on Local Differential Privacy (LDP) mechanisms and formalizing the problem statement. Section \ref{section:model} provides a detailed description of our proposed method, \emph{CLDP}. Next, Section \ref{section:setting} presents the experimental setup, including implementation details and the datasets used to evaluate our model. Section \ref{section:result} analyzes and discusses the key findings from our experiments. Section \ref{section:related} surveys recent advances in Differential Privacy (DP) and LDP for time series data, contextualizing our work within the broader literature. Finally, Section \ref{section:conclusion} summarizes the key contributions of this study, outlines future research directions, and concludes the paper.

\section{Preliminaries and Problem Statement}\label{section:preli}

\subsection{Preliminaries: Local Differential Privacy}

Local Differential Privacy (LDP) is a method for preserving privacy by perturbing data at the individual level before aggregation, ensuring that original data cannot be inferred with high confidence. Mathematically, LDP is defined as follows \cite{arachchige2019local}:

\[
Pr[\mathcal{A}(x) = y] \leq e^{\epsilon} \times Pr[\mathcal{A}(x') = y]
\]

where \( \epsilon \) controls the privacy level, with smaller values indicating stronger privacy. The main perturbation mechanisms in LDP are the Laplace mechanism and Generalized Randomized Response. For time series data, LDP approaches are divided into value perturbation (VLDP) and temporal perturbation models (TLDP). In TLDP, neighboring time series are defined as those that can be converted by swapping values at two specific timestamps. The concept of \( \epsilon \)-TLDP ensures that the privacy condition holds across neighboring time series within a defined time window.

In contrast, VLDP mechanisms add noise directly to the values in the series to preserve privacy, while maintaining the statistical properties. Common methods include the Laplace and Gaussian mechanisms. These approaches provide a trade-off between privacy protection and data utility. Each of these mechanisms offers different privacy guarantees based on noise distribution and parameters, providing foundational techniques for privacy-preserving time series analysis.

\subsection{Problem Definition}

Traditional LDP-based methods for time series data perturb each user's data point individually, introducing privacy vulnerabilities when the data collector is malicious. For example, given a time series \(X = \{x_1, x_2, \dots, x_n\}\), standard approaches generate a perturbed series \(\tilde{X} = \{x_1 + \eta_1, x_2 + \eta_2, \dots, x_n + \eta_n\}\), where each \(\eta_i\) is independently added noise. However, when aggregative operations such as summation or averaging are performed over fixed time windows, the noise can partially or fully cancel out, enabling adversaries to approximate original values. Moreover, these methods overlook the temporal correlations in time series data, increasing susceptibility to inference attacks.

To mitigate these limitations, we propose a \emph{Cooperative LDP (CLDP)} method that enhances privacy by distributing noise vectors across multiple users. Instead of each user independently adding noise, a correlated noise vector space is collaboratively generated and shared among \(n\) users. Each user then generates noise within this space and perturbs their data with a designated noise component before transmitting it to the data collector. During aggregation (e.g., summation), the perturbation noise is structured to cancel out, preserving only the aggregated statistical values while protecting individual privacy. This approach strengthens privacy guarantees without significantly compromising data utility, providing a more robust defense against inference attacks in time series data.

%\vspace{-0cm}
\hspace{0cm}
\begin{table}[t]
\small
\caption{Systems' parameters}
\vspace{-6pt}
\begin{center}
\begin{tabular}{|c|c|}
\hline

\textbf{Parameters}&{\textbf{Symbols}} \\
\hline

\text{Time series data} & {$X$} \\
\hline
\text{Tossing space}&{\text{$k$}} \\
\hline

\text{Number of users}&{\text{$u$}} \\
\hline
\text{Window/Partition size of data sample} & {$l$} \\
\hline

\text{Noise amplitude}&{A} \\
\hline
\text{Time cycle}&{T} \\
\hline
\text{Privacy budget}&{$\epsilon$} \\
\hline
\text{Perturbed noise }&{$\lambda$} \\
\hline
\end{tabular}
\label{tab1:parameters}
\end{center}
\end{table}

%Traditional LDP-based methods for time series data typically perturb each user’s data point individually. For example, given a time series \(X = \{x_1, x_2, \dots, x_n\}\), the standard approach produces a perturbed series \(\tilde{X} = \{x_1 + \eta_1, x_2 + \eta_2, \dots, x_n + \eta_n\}\), where each \(\eta_i\) represents the noise added to \(x_i\). However, this strategy introduces significant privacy gaps when the data collector is malicious. When noise is applied over fixed time windows—say, for a window \(T\)—aggregative operations such as summation, $S_T = \sum_{i \in T} (x_i + \eta_i),$ or averaging, $\bar{x}_T = \frac{1}{|T|} \sum_{i \in T} (x_i + \eta_i),$ can inadvertently cancel out the injected noise. This cancellation allows adversaries (i.e., data collector) to approximate the original statistical value of individual. Moreover, these methods often disregard the temporal correlations inherent in time series data, thereby facilitating inference attacks over extended periods. Consequently, there is a pressing need for innovative privacy-preserving mechanisms that consider both the cumulative effects of noise and the sequential nature of the data, ensuring robust protection against privacy breaches without significantly compromising utility.
\vspace{-4pt}
\section{Proposed Method: Cooperative LDP}\label{section:model}

\subsection{Motivation}
This research is motivated by the shortcomings of traditional LDP methods for time series data. For example, consider power consumption data collected from smart meters in a residential area. In traditional LDP approaches, each household's power usage data is perturbed independently by adding random noise. However, when these noisy data points are aggregated—such as calculating the total energy demand over a specific time period—the noise can cancel out, allowing an adversary to closely estimate individual consumption patterns. This vulnerability not only risks revealing household activities (e.g., when residents are home or using appliances) but also ignores the temporal correlations in power usage, making the system more susceptible to inference attacks. Our work addresses these issues by proposing a cooperative noise distribution approach, where noise is collaboratively generated and assigned among multiple users. This ensures that, upon aggregation, privacy is preserved while maintaining data utility, making the system more secure and effective for large-scale, real-time energy monitoring applications.

\subsection{Technical Details of Cooperative LDP}

\begin{figure*}[t]
    \centering
    \includegraphics[width=0.90\textwidth]{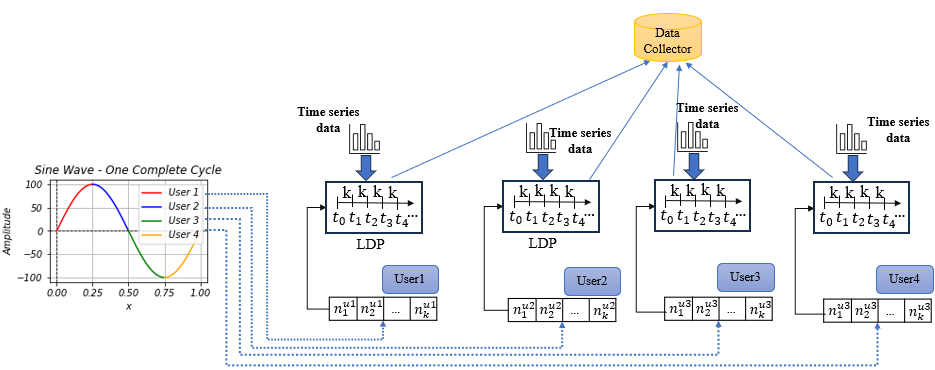}
    \caption{Overall Architecture for Cooperative LDP}
    \label{fig:architecture}
\end{figure*}

Our proposed Cooperative Local Differential Privacy (LDP) mechanism utilizes a complete sine wave to generate noise for perturbing time series data. This noise is designed to cancel out during aggregation at the data collector, thereby preserving data utility while ensuring privacy. Figure \ref{fig:architecture} illustrates the overall architecture of our approach. To explain this process mathematically, we define key components and steps as follows.

{\noindent{\bf{Partitioning the Sine wave.}}} To distribute noise effectively, we divide a complete sine wave cycle (amplitude scaled as needed) into 
$u$ equal symmetric partitions, where $u$ represents the number of users. Each partition corresponds to the noise assigned to a single user, ensuring that the noise contributions from all users collectively cancel out when aggregated. Suppose that the time series dataset $X$ of length  $T$ be divided into non-overlapping time windows. Mathematically, a complete sine wave is represented as: $  y(t) = A \cdot \sin\left(\frac{2\pi t}{T}\right)$,  where \( A \) is amplitude, \( T \) is the period of the sine wave, \( t \) is time variable. To divide the sine wave into equal segments for $u$ users, we can write this for $i$ th user is: $   \ t_i= \left[\frac{(i-1) \cdot T}{u}, \frac{i \cdot T}{u}\right), $ for \( i = 1, 2, \ldots, u \). This ensures the sine wave is split uniquely among \( u \) users without overlap.

Now, we assign the number of tosses $k$ for each partition. Each user’s partition has an equal tossing space distributed uniformly within their assigned interval. For a partition with interval length \( \Delta T = \frac{T}{u} \), the tossing points would be: 
\[ \left\{ t_{i,j} = \frac{(i-1) \cdot T}{u} + j \cdot \frac{\Delta T}{k} \, \Big| \, j = 0, 1, \ldots, k-1 \right\}.
\]  
For instance, if \( k = 6 \) and the partition is \( [0, 0.25] \), the tossing points are:  
$\{0, 0.05, 0.1, 0.15, 0.2, 0.25\}.$  This ensures the noise is spread uniformly across the partition.

Moreover, the number of tossing times is determined by the number of data samples \(l\) that need to be perturbed. In this setup, \(k\) can either be equal to or less than \(l\), meaning that \(k\) defines the level of granularity for perturbing the time series data for a given user. Each user will have an equal number of data samples for perturbation. After obtaining the noise values (i.e., the sine wave amplitudes at these tossing points), we shuffle them and add the shuffled noise to the original time series data. The number of data samples $l$ in each time window determines how many times the noise is applied within that window. By choosing $k$ to be small relative to $l$, the probability that the same noise values are used across the samples is high, leading to effective noise cancellation during aggregation. This, in turn, preserves the overall accuracy of the aggregated data while ensuring individual privacy.

{\bf{\noindent{Amplitude as Noise.}}} Each tossing point corresponds to a amplitude value on the sine wave, which is then considered as the noise for that particular sample. The amplitude $A$ of the sine wave determines the magnitude of the noise, and the noise value is calculated as:
$$
y_i(t_i) = 
\begin{cases} 
A \cdot \sin\left(\frac{2\pi t_i}{T}\right), & t_i \in \left[\frac{(i-1) \cdot T}{u}, \frac{i \cdot T}{u}\right), \\
0, & \text{otherwise}.
\end{cases}
$$

This noise is then applied to the time series data, perturbing the original values by adding or subtracting these noise values. After calculating the noise for each tossing point, the amplitudes corresponding to the tossing points are shuffled. This shuffling step is crucial to ensure that the noise is not correlated in any predictable way. The shuffled noise values are then added to the original time series data, providing perturbation and thus ensuring privacy.
\begin{figure}[t]
	\centering
	\captionsetup{justification=centering}
	\includegraphics[width=0.9\linewidth]{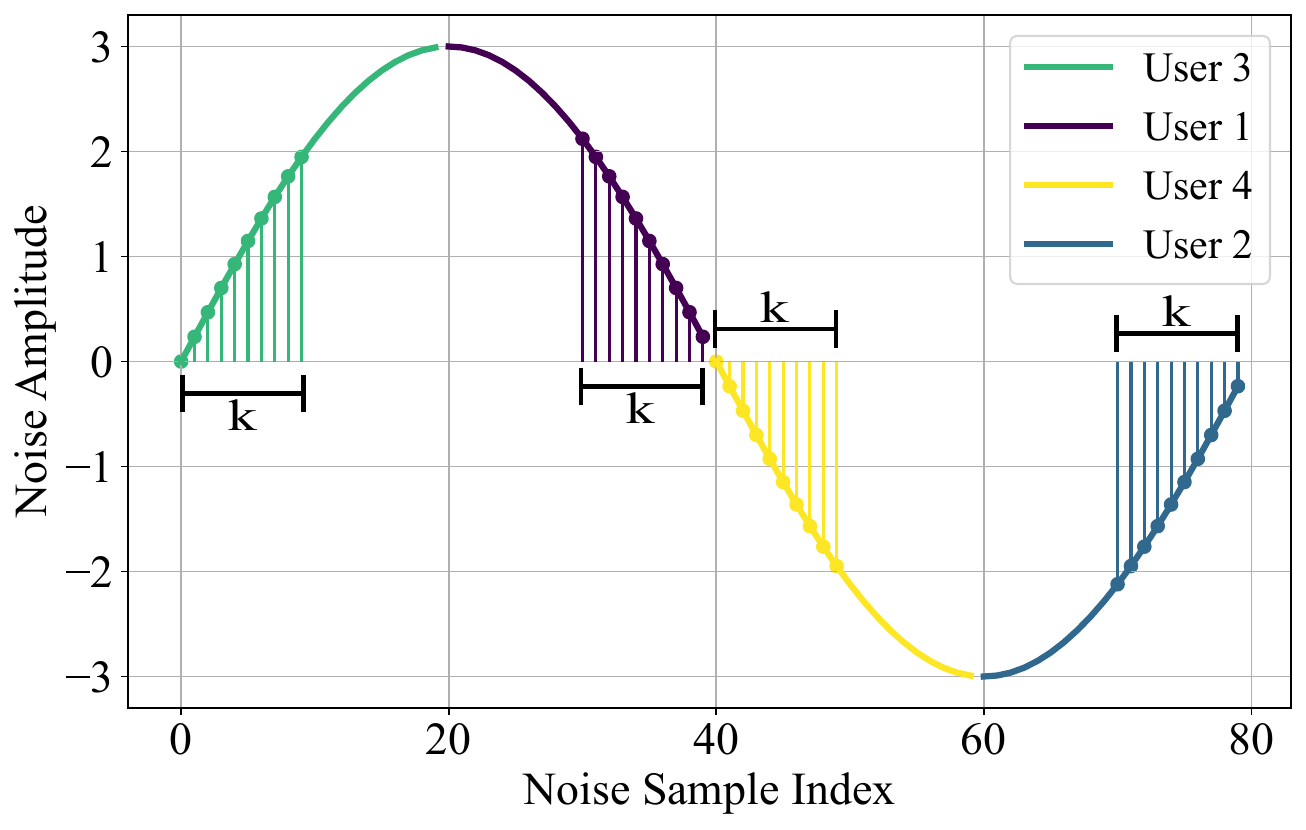}
	\caption{Sine Wave Partitioning And Noise Distribution}
	\label{fig:sine_wave_partitioning}
	\vspace{-1.3\baselineskip}
\end{figure}

The figure \ref{fig:sine_wave_partitioning} visually represents how noise is structured and distributed among multiple users in our Cooperative Local Differential Privacy (CLDP) framework using a sine wave-based noise generation approach. Additionally, Table \ref{tab1:parameters} provides a comprehensive list of parameters along with their corresponding symbols used throughout the study.

{\bf{\noindent{Trade-offs between Privacy and Utility.}}} In our approach, the privacy budget is determined by how noise is applied over the time series data, specifically by the number of tossing points \( k \) and the number of samples \( l \) in each time window. Here, \( k \) represents the number of evenly spaced tossing points within each partition of the sine wave, while \( l \) denotes the number of data samples in a given time window that are perturbed using these noise values. When \( k \) is small relative to \( l \), the same noise value is likely to be applied across many samples, which promotes noise cancellation during aggregation and thereby enhances data utility. At the same time, this repeated application—if properly randomized via shuffling—limits the adversary’s ability to infer the original data, thus preserving privacy. Mathematically, a simplified representation of the privacy budget might be expressed as:

\[
\epsilon \propto \frac{l \times u }{A \times k}
\]

In this relationship, \(l\) represents the number of data samples in a given time window, and \(u\) is the number of users. These factors in the numerator imply that an increase in either the sample size or the number of users leads to a higher privacy loss (i.e., a larger \(\epsilon\)), potentially weakening privacy protection. Conversely, \(A\) denotes the amplitude of the sine wave and \(k\) represents the tossing space; both of these parameters in the denominator contribute to noise strength. Increasing \(A\) or \(k\) enhances the magnitude and distribution of the noise, thereby reducing \(\epsilon\) and strengthening privacy. Balancing these parameters is crucial for achieving strong privacy guarantees without sacrificing the utility of the aggregated time series data.

\subsection{Security Analysis}

The probability of breaking the security of perturbed values in a system where \( k \) represents the tossing space and \( u \times l \) is the total number of tosses depends on the uniform distribution of values over the tossing space. The mechanism assumes that each toss is independent and uniformly random. This helps quantify the adversary's chance of reconstructing the original values based on the perturbed data.

The probability of an adversary successfully guessing or reconstructing all perturbed values is expressed as:

\[
P_{\text{break}} = \left(\frac{1}{k}\right)^{l \times u}
\]

where, \( k \) is the tossing space, or the range of possible values for each toss, \( l \) is the number of data samples being perturbed, \( u \) is the number of users involved in the perturbation process, \( l \cdot u \) represents the total number of tosses performed.

The security of the perturbation mechanism improves with the parameters \( k \), \( l \), and \( u \). Increasing \( k \) (the size of the tossing space) exponentially reduces the probability of successful guessing, as larger spaces make each guess more uncertain. Similarly, increasing \( l \) (number of data samples) or \( u \) (number of users) enhances security because it increases the total number of independent tosses, making the probability of breaking security even smaller. For example, if \( k = 10 \), \( l = 5 \), and \( u = 3 \), the breaking probability becomes \( P_{\text{break}} = 10^{-15} \), which demonstrates extremely high security.

This model relies on a perfectly uniform distribution to ensure robustness. Any deviations or correlations between tosses may weaken security, enabling adversaries to exploit patterns. While increasing \( k \), \( l \), and \( u \) strengthens protection, careful design of the tossing mechanism is crucial to maintain randomness, balance privacy, and preserve data utility.

\subsection{Proposed Algorithm}
The suggested algorithm disturbs data by introducing structured noise while ensuring symmetry. It divides a complete sine wave among users with an even distribution. Users are randomized, and noise partitions are assigned iteratively. Each partition has a specified noise space, and values are chosen based on quadrant location. The chosen noise values are then used to disturb data points. This is repeated until all users have been assigned their noise partitions, ensuring randomized and balanced disturbance.

The algorithm \ref{alg:proposed} implements a Custom Local Differential Privacy (CLDP) method using a sine wave to generate distributed noise for multiple users. It starts by dividing a full sine wave equally among users (\textbf{Step 1}), ensuring that the number of users is a multiple of 2 or 4 for symmetric noise distribution. For each user (\textbf{Step 2}), noise samples $\lambda$ are computed based on the given amplitude $A$ and tossing space $k$. These noise samples are then randomly shuffled (\textbf{Step 3}) to remove direct correlations. Finally, the shuffled noise values are added to the original data points (\textbf{Step 4}), producing the perturbed data stream $l'$. The process iterates over all users (\textbf{Step 5}) until all data samples are perturbed, ensuring privacy preservation while maintaining data utility. Finally, when the collector aggregates all the perturbed data samples from all users, the added noise will nearly cancel out, preserving the overall data integrity.

\begin{algorithm}[h]
    \caption{CLDP using Sine Wave for Distributed Noise Generation}
    \label{alg:proposed}
    \KwInput{Data sample size ($l$), tossing space ($k$), number of users ($u$), noise amplitude ($A$)}
    \KwOutput{Perturbed Data Stream $l'$}
    
    \myproc{NoisePerturbation()}{
        \tcp{Step 1: Generate Noise Samples}
        Divide a full sine wave into $u$ equal partitions, ensuring $u$ is a multiple of $2$ or $4$ for symmetric noise distribution\;
        
        \ForEach{user $i \in u$}{
            \tcp{Step 2: Compute Noise Space}
            Generate noise samples $\lambda$ based on the amplitude $A$ and tossing space $k$\;

            \tcp{Step 3: Shuffle Noise Samples}
            Randomly shuffle the noise values in $\lambda$\;
            
            \tcp{Step 4: Apply Noise to Data}
            Perturb each data sample by adding its corresponding noise value: $l'_{i} = l_{i} + \lambda_{i}$\;
        }
        \tcp{Step 5: Repeat for All Users}
    }
\end{algorithm}

\section{Experimental Setting and Dataset}  \label{section:setting}

\subsection{Dataset}

We have conducted our experiments on both real-world and synthetic time series data to evaluate the performance of our Cooperative Local Differential Privacy (CLDP) strategy.

\noindent {\bf{{Real-World Dataset (Individual Household Electric Power Consumption{\footnote{\url{https://archive.ics.uci.edu/dataset/235/individual+household+electric+power+consumption}}}).}}} We employed the Individual Household Electric Power Consumption data set available in the UCI Machine Learning Repository. The data set contains detailed time series household electric consumption data, and thus it is a suitable candidate to evaluate privacy-preserving mechanisms for real-world applications. Since our approach is aimed at multi-user data collection with Local Differential Privacy (LDP), we split the data set into four segments, where each segment represents a single user. This setup allows us to demonstrate how our CLDP approach ensures privacy preservation with data utility in a real-world scenario.

\noindent {\bf{Synthetic Dataset.}}
In addition to real data, we also generated a synthetic time series dataset to study the privacy-utility trade-off for multiple users under different privacy budgets. The synthetic dataset includes a number of users' sequential data, where each user contributes time series records. In this artificial setting, we can examine systematically the impact of different levels of noise perturbation and aggregation approaches on the overall utility of the dataset.

On real and synthetic datasets, our experiments provide a comprehensive assessment of the data utility and privacy guarantee of the proposed CLDP approach, confirming its effectiveness under diverse settings.

\subsection{Setting}

To determine the effectiveness of our proposed Cooperative Local Differential Privacy (CLDP) method, we performed our experiments using Google Colab in its default configuration. Google Colab{\footnote{\url{https://colab.research.google.com/drive/159ReE5crixpRqO6FDWdhSUWBXxHDLegv}}} is an online Python coding environment that supports native integration with essential libraries used for data processing, privacy protection, and statistical analysis. We used Python as the primary programming language and libraries such as NumPy, Pandas, and Matplotlib to manipulate and visualize data. The experiments were executed on Google Colab's standard runtime, utilizing a cloud-based CPU and GPU where available

\section{Experimental Evaluation}\label{section:result}

We evaluated the effectiveness of our \emph{CLDP} approach through a series of experiments designed around four key objectives: i) Assessing Data Distortion: we measured the distortion introduced to the original time series data after applying perturbations; ii) Varying tossing size: we computed the Mean Squared Error (MSE) by varying the tossing space parameter ($k$) and the number of users ($u$) while keeping the window size of data samples ($l$) and the noise amplitude ($A$) constant; iii) Altering the Window Size: we analyzed the MSE when adjusting the window size ($l$) in conjunction with $A$ from the set ($k$, $u$, $A$), while keeping the other two parameters constant; iv) Changing Noise Amplitude: Finally, we evaluated the MSE by modifying the noise amplitude ($A$) alongside $u$ from ($k$, $u$, $l$), with the other two parameters maintained at fixed values.

This systematic exploration enabled us to isolate and understand the influence of each parameter on the performance of our CLDP approach.

\vspace{-4pt}
\subsection{Original Data Distortion of Multiple Users}
\begin{figure}[t]
	\centering
	\captionsetup{justification=centering}
	\includegraphics[width=0.9\linewidth]{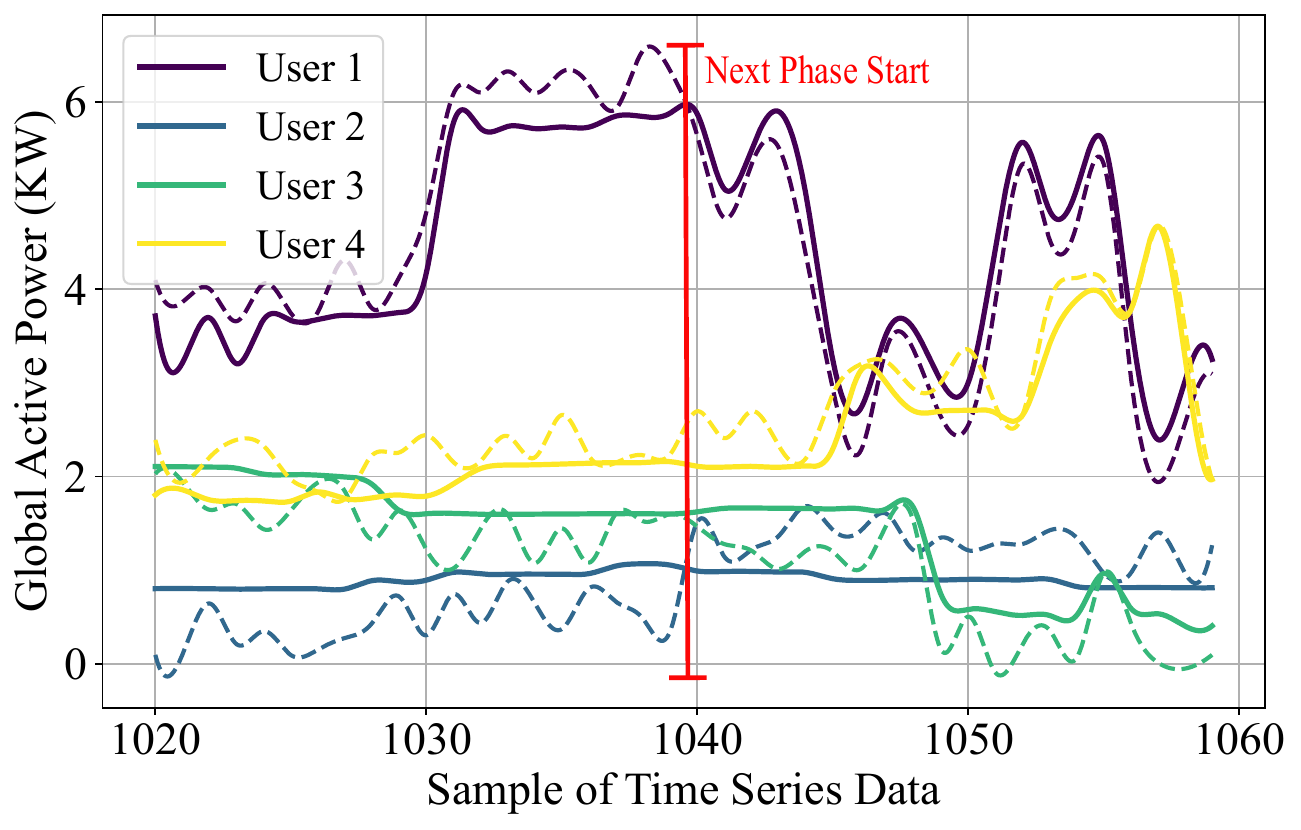}
	\caption{Perturbed VS Original Data}
	\label{fig:original_vs_perturbed}
	\vspace{-1.3\baselineskip}
\end{figure}

\vspace{-4pt}
In order to estimate the impact of our Cooperative Local Differential Privacy (CLDP) mechanism on data distortion, we compared original and perturbed time series data across different users, see figure \ref{fig:original_vs_perturbed}.

The graph illustrates fluctuations in Global Active Power (KW) readings from four users when introducing noise perturbation under the LDP protocol. The x-axis is a sample segment of the time series data, and the y-axis indicates power consumption levels. Each user's raw data was perturbed before being aggregated to guarantee privacy at the individual level.

As observed in the plot, despite the disrupted values having randomness after adding noise, the trend in the overall data remains unaffected. This is depicted to demonstrate the effectiveness of our noise generation paradigm in maintaining underlying characteristics of the dataset without weakening on leakage of sensitive data. In addition, by providing distributed noise vectors to users, aggregated results in entirety can prove better than what normal LDP methodology generally suffers with.

In our second evaluation phase, we analyze the effect of changing privacy budgets on this distortion with a tradeoff between data utility and privacy protection.

\subsection{MSE for Different Size of Tossing Space}
To analyze the privacy-utility trade-off of our Cooperative Local Differential Privacy (CLDP) solution, we analyzed how varying the size of the tossing space (k) affects the added noise, in terms of Mean Squared Error (MSE). Since MSE quantifies the noise, it is directly proportional to the privacy budget—a larger MSE means a stronger privacy guarantee but at the cost of utility.

\begin{figure}[t]
	\centering
	\captionsetup{justification=centering}
	\includegraphics[width=0.9\linewidth]{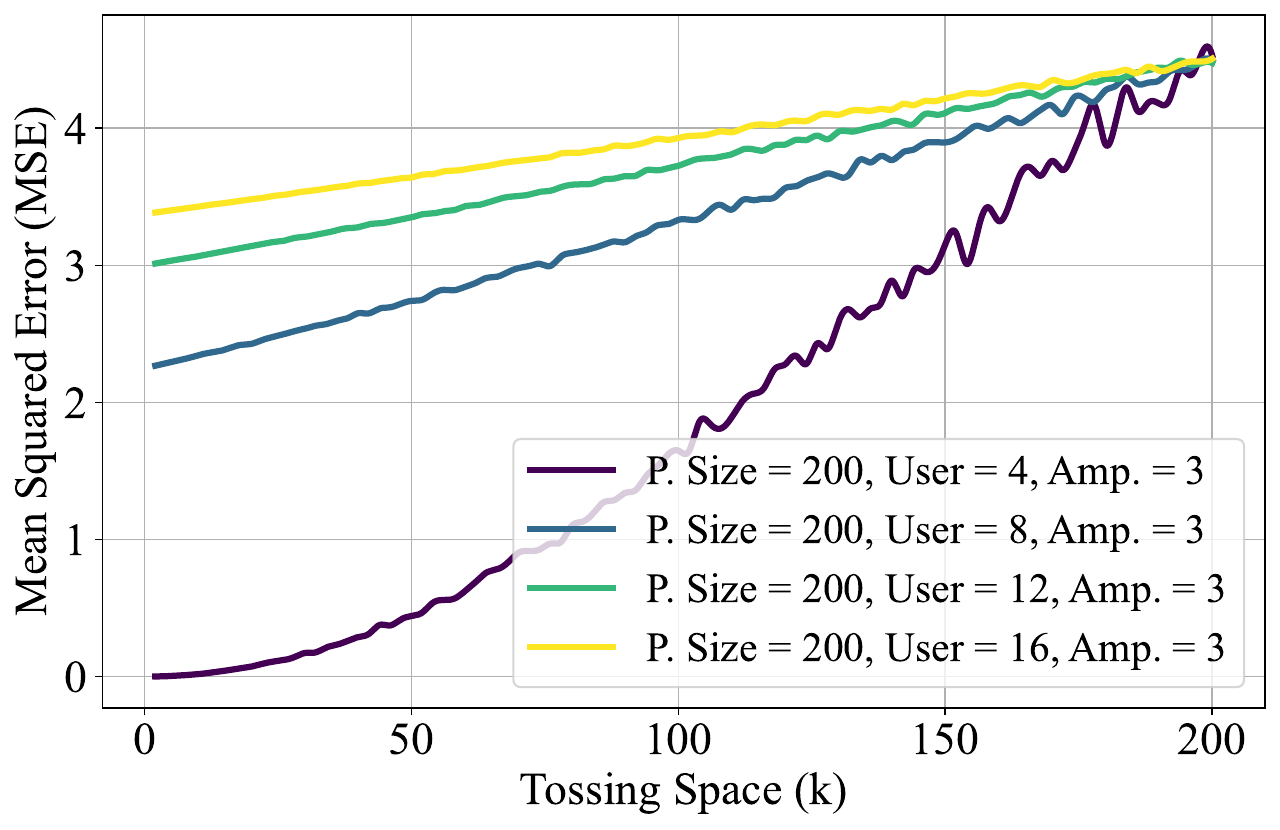}
	\caption{MSE for Varying Tossing Space ($k$) vs. Numbers of Users ($u$) }
	\label{fig:mse_vs_k}
	\vspace{-1.3\baselineskip}
\end{figure}

In figure \ref{fig:mse_vs_k} the graph plots the variation of MSE with the tossing space (k) for different user settings. We considered a number of settings where the number of users was 4, 8, 12, and 16, and the partition size and noise amplitude from sine wave were fixed at 200 and 3, respectively. The x-axis has the size of the tossing space, and the y-axis the corresponding MSE. Key observations from the results:
\begin{enumerate}
\item {As tossing space (k) increases, MSE also increases, which means the addition of more noise and thus greater protection of privacy.}
\item {The impact of tossing space on MSE is greater for larger user groups since the cooperative noise distribution method achieves a collective cumulative perturbation effect.}
\item {Despite the added noise, the structured noise generation approach ensures that aggregated values retain useful patterns in them.}
    
\end{enumerate}

The experiments illustrate the value of adaptive tossing techniques for fine-tuning the privacy-data usefulness trade-off, demonstrating how different configurations can be calibrated based on specific privacy requirements.

\subsection{MSE for Different Size of Data Sample Window}
This experiment analyzes the influence of Partition Size (l) on the injected noise as per Mean Squared Error (MSE) in varying user configurations. In our Cooperative Local Differential Privacy (CLDP) framework, partition size determines the amount of granularity with which the data is segmented, which directly impacts the distribution of noise injected into the time series data.

\begin{figure}[t]
	\centering
	\captionsetup{justification=centering}
	\includegraphics[width=0.9\linewidth]{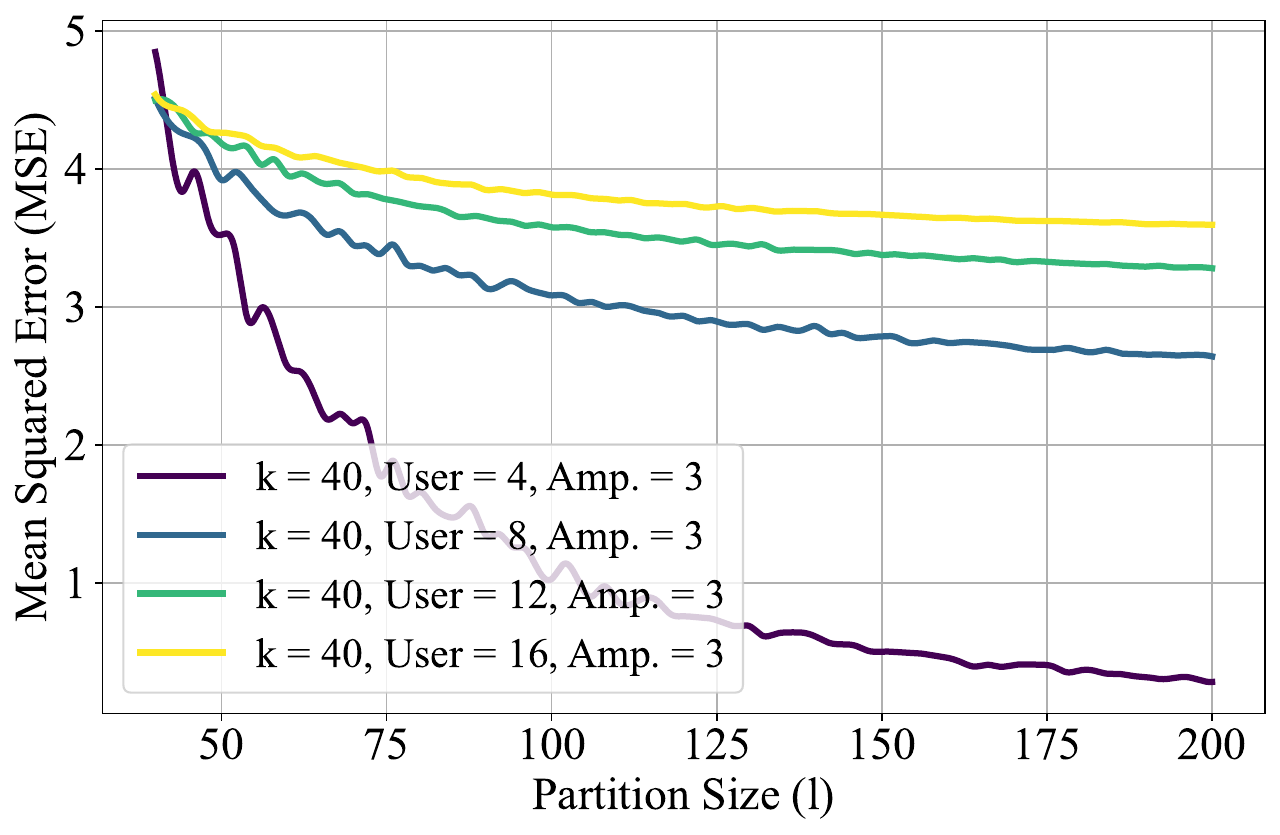}
	\caption{MSE for Varying Data Sample ($l$) vs. Numbers of Users ($u$)}
	\label{fig:mse_vs_l}
	\vspace{-1.3\baselineskip}
\end{figure}

The graph in figure \ref{fig:mse_vs_l} indicates the influence of Partition Size (l) on MSE for user sizes 4, 8, 12, and 16. Surprisingly, when the partition size increases, MSE decreases. This is because a larger partition size has the implication of squeezing the partitions vertically. With the same tossing space and amplitude, the sine wave noise gets distributed over a greater section of data, resulting in less vertical spread and hence total noise value (i.e., smaller MSE). Growing partition size distributes the same noise pattern over a wider range, resulting in vertical compression of the noise space. Since k is an amplitude-based quantity and is held constant, growing l amounts to less per-segment fluctuation of noise and thus a smaller MSE.

\vspace{-4pt}
\subsection{MSE for Varying the Number of Users}
\begin{figure}[t]
	\centering
	\captionsetup{justification=centering}
	\includegraphics[width=0.9\linewidth]{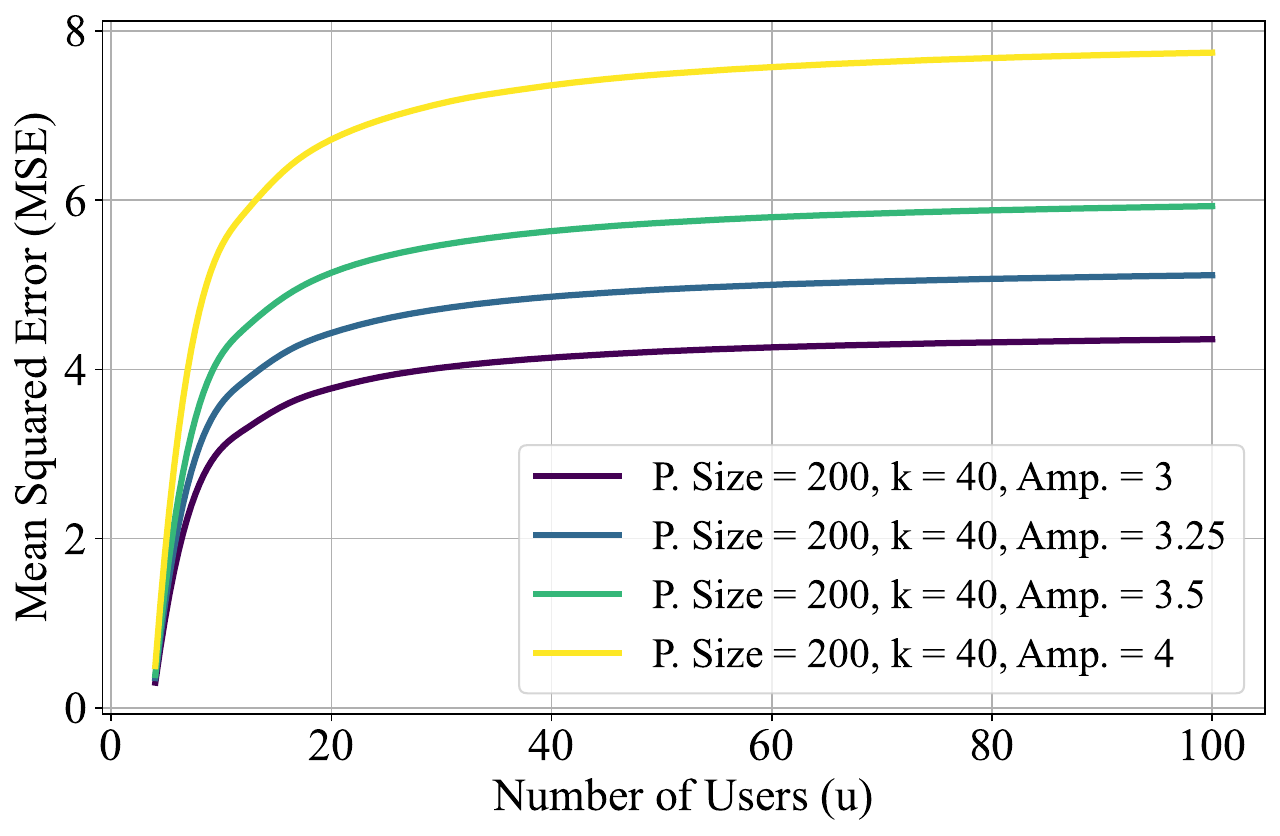}
	\caption{MSE for Varying Number of Users vs. Amplitude (A)}
	\label{fig:mse_vs_user}
	\vspace{-1.3\baselineskip}
\end{figure}

\vspace{-4pt}
In this experiment, we investigate how the Mean Squared Error (MSE) varies as the number of users (u) is varied in our Cooperative Local Differential Privacy (CLDP) system. The experiment was conducted with a fixed Partition Size (P. Size = 200) and Tossing Space (k = 40), but varying Sine Wave Amplitude (3, 3.25, 3.5, and 4). The findings reflect the following significant trends in figure \ref{fig:mse_vs_user}:
\begin{enumerate}
\item {As the number of users increases, the MSE grows at an early stage very rapidly, as would be expected, since more users contribute independently perturbed values that contribute to the overall noise in the system.}
\item {After some point in terms of the number of users, the growth of MSE will be less, i.e., more users contribute less and less to the noise overall. This means that the additional noise from additional users starts plateauing.}
\item {Larger sine wave amplitudes yield uniformly higher MSE values for all user numbers. The overall trend, however, remains the same—a rapid rise initially and then an accelerating rate.}
    
\end{enumerate}

This is an important observation since it points out that although adding more users improves privacy protection, beyond a point, the added noise settles down such that data utility is not too much affected. This property of CLDP enables scalability without severely degrading the quality of the data beyond a point.

\subsection{MSE for Different Size of Amplitude}
\begin{figure}[t]
	\centering
	\captionsetup{justification=centering}
	\includegraphics[width=0.9\linewidth]{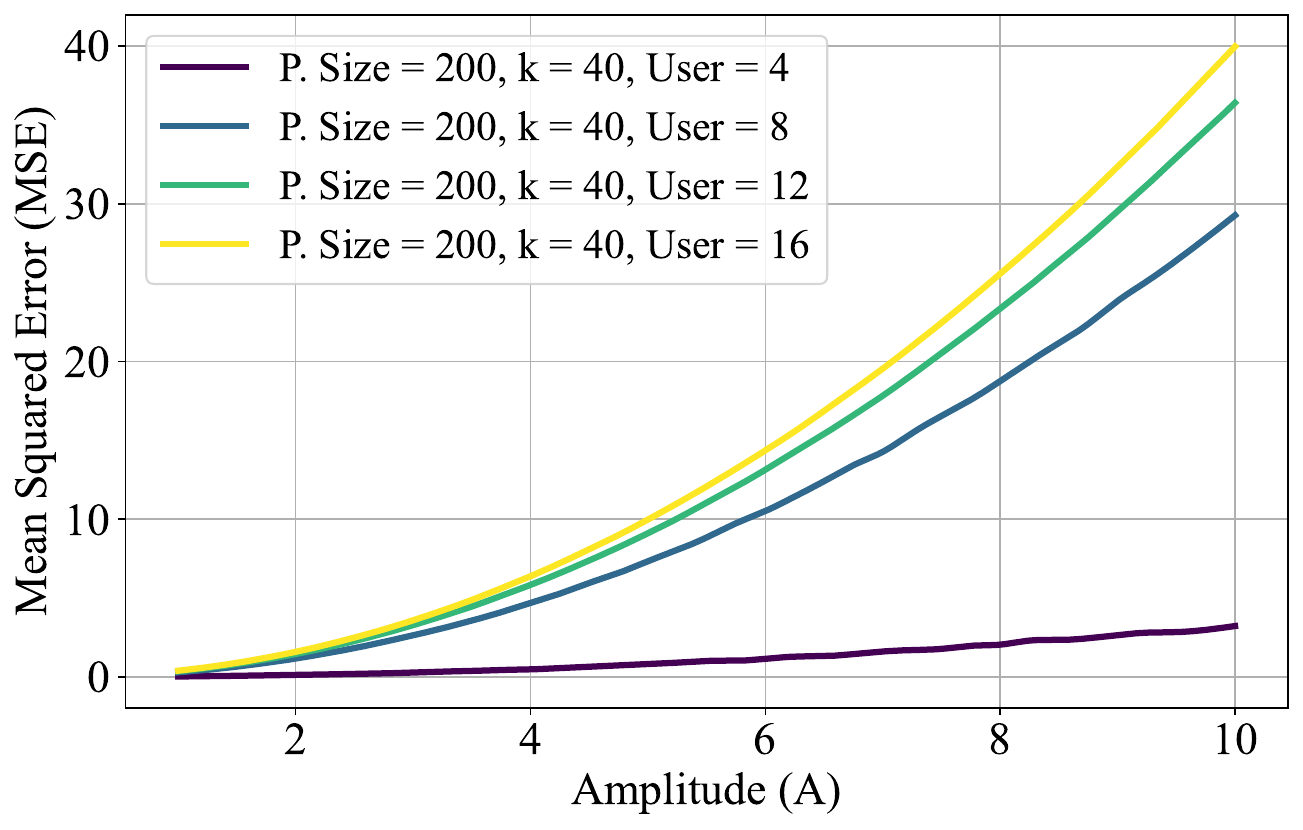}
	\caption{ MSE for Varying Amplitude (A) vs. Number of Users.}
	\label{fig:mse_vs_A}
	\vspace{-1.3\baselineskip}
\end{figure}

Amplitude directly affects the noise space as it controls the range of values in which perturbations occur. The experiment was conducted with fixed Partition Size (P. Size = 200) of data samples, Tossing Space ($k = 40$), and varying user numbers ($4, 8, 12,$ and $16$). In this experiment, we analyze the effect of varying sine wave amplitude on the Mean Squared Error (MSE) in our Cooperative Local Differential Privacy (CLDP) mechanism.
The graph illustrates in figure \ref{fig:mse_vs_A} how MSE changes with the increasing amplitude for a changing number of users. As the amplitude increases, the MSE also increases significantly. This is expected because with an increasing amplitude comes a larger noise space, which provides greater privacy protection but greater data distortion. Since the noise space is determined by the amplitude of the sine wave, greater amplitude indeed enlarges the scope of perturbation, and hence the overall data perturbation becomes stronger, particularly when multiple users are involved in the aggregation.

This experiment illustrates the crucial role played by amplitude tuning in balancing privacy and data utility. A well-chosen amplitude achieves satisfactory privacy protection while maintaining sufficient data accuracy, and therefore is a primary parameter in real-world CLDP deployments.

\section{Related Work}\label{section:related}
Privacy concerns surrounding time series data have gained significant attention due to their widespread use in various applications, such as wearable health monitoring and smart home systems. The continuous nature of time-series data makes them susceptible to privacy breaches, as periodic patterns, such as heart rate fluctuations or power consumption, can inadvertently reveal sensitive user behaviors \cite{allard2015chiaroscuro, zheng2021efficient, dwork2006differential}. To mitigate these risks,  existing LDP-based methods for time series data often struggle with balancing privacy and utility. Some techniques selectively perturb data points to preserve overall patterns, while others prioritize privacy through randomized perturbation, which can obscure meaningful trends. Prior research has demonstrated that LDP-based privacy mechanisms may degrade data utility, making it difficult to extract meaningful insights from perturbed datasets \cite{bittau2017prochlo, qin2016heavy}. To address these trade-offs, researchers have proposed alternative strategies, such as partitioning time series data into time windows and selectively perturbing specific samples to retain structural integrity \cite{fahrenkrog2024privacy, gao2023privacy, katsomallos2022landmark}. However, these methods often suffer from scalability issues and computational overhead, particularly in high-frequency or real-time scenarios. Additionally, preserving inherent data patterns may inadvertently expose user behaviors. For instance, in smart home energy monitoring, maintaining temporal patterns could reveal occupancy patterns, posing a privacy risk.  

To tackle these challenges, we propose a novel privacy-preserving mechanism, \emph{Cooperative LDP}, which leverages structured noise generation based on sine wave properties. Unlike conventional LDP methods that introduce independent noise per user, our approach partitions a sine wave among multiple users, ensuring that individual perturbations cancel out when aggregated. This structured noise distribution allows for strong privacy guarantees while minimizing distortions in the collected data. Furthermore, our method incorporates an adaptive tossing strategy, enabling fine-grained control over privacy budgets and noise distribution. By optimizing perturbation placement and noise allocation, Cooperative LDP effectively balances privacy preservation and analytical utility, facilitating privacy-aware time series data sharing with minimal loss of information.
\vspace{-1pt}
\section{Conclusion}\label{section:conclusion}
\vspace{-4pt}
In this work, we introduced a Cooperative Local Differential Privacy (CLDP) mechanism that enhances privacy while maintaining the analytical utility of time series data. By leveraging structured noise generation based on a sine wave, our method ensures that individual perturbations cancel out during aggregation, preserving statistical characteristics while safeguarding user privacy. The adaptive tossing strategy allows users to fine-tune their privacy budgets, striking a balance between data protection and usability. Our security analysis demonstrates that as the number of users and data samples increases, adversarial reconstruction of original data becomes exponentially more challenging. This approach offers a scalable and effective privacy-preserving solution for time series data in real-world applications such as smart homes, healthcare monitoring, and financial analytics.  
 
Despite its advantages, our CLDP mechanism has some limitations. First, the method relies on the assumption that the number of participating users remains stable; significant fluctuations in user participation could disrupt the noise cancellation effect. Additionally, our approach assumes that data is collected in structured time windows, which may not always align with real-world scenarios where data streams are irregular or asynchronous. Future work should explore adaptive noise strategies that dynamically adjust to changing data distributions and user participation levels to further improve privacy guarantees.

\section*{Acknowledgments}
This work is supported by Coastal Virginia node of the Commonwealth Cyber Initiative (COVACCI).

%\begin{thebibliography}{1}

\bibliographystyle{IEEEtran}
\bibliography{IEEEexample}

%\end{thebibliography}
\end{document}